\documentclass{emulateapj}
\usepackage{apjfonts}
\usepackage{graphicx}
\usepackage{epsfig}
\usepackage{rotating}
\pdfoutput=1
\makeatletter

\newcommand{\bc}{\begin{center}}
\newcommand{\ec}{\end{center}}
\newcommand{\be}{\begin{equation}}
\newcommand{\ee}{\end{equation}}
\newcommand{\ba}{\begin{eqnarray}}
\newcommand{\ea}{\end{eqnarray}}
\newcommand{\bt}{\begin{tabular}}
\newcommand{\et}{\end{tabular}}

\begin{document}

\title{Young energetic PSR J1617--5055, its underluminous nebula and unidentified TeV source HESS J1616--508}

\author{O.\  Kargaltsev\altaffilmark{1,2},   G.\ G.\
Pavlov\altaffilmark{2}, and J.\ A.\ Wong\altaffilmark{2}}

\altaffiltext{1}{ Dept.\ of Astronomy, University of Florida, Bryant
Space Science Center, Gainesville, FL 32611; oyk100@astro.ufl.edu}

\altaffiltext{2}{The Pennsylvania State University, 525 Davey Lab,
University Park, PA 16802, USA; pavlov@astro.psu.edu}

\begin{abstract}

 We observed the young energetic pulsar J1617$-$5055
  with the {\sl Chandra} ACIS detector for 60 ks.
In addition to the pulsar, the X-ray images
show a faint  pulsar-wind nebula (PWN)
seen up to $\sim1'$ from the pulsar.
Deconvolution and
reconstruction of the image reveal
a brighter
 compact PWN component
of $\sim1''$ size, possibly with a jet-torus morphology. The  PWN
spectrum fits an absorbed power-law (PL)
 model with
 the photon index $\Gamma\approx1.5$. The total PWN luminosity in the 0.5--8 keV band, $L_{\rm
pwn}=3.2\times 10^{33}$ ergs s$^{-1}$ for $d=6.5$ kpc, is a fraction
of $2\times 10^{-4}$ of the pulsar's spin-down power $\dot{E}$ and
a fraction of 0.2 of the pulsar's X-ray luminosity, which is a factor of
20 lower than one would expect
 from an average empirical relation
found for a sample
 of PWNe observed with {\sl Chandra}.
The pulsar's spectrum can be described by an absorbed PL with
$n_{\rm H}\approx 3.5\times10^{22}$ cm$^{-2}$ and
$\Gamma\approx1.1$, harder than any other pulsar spectrum
reliably measured in the soft X-ray band. This non-thermal emission
is $\approx50\%$ pulsed showing one peak per period.
  We have also investigated a possible connection between J1617 and the
extended TeV source HESS J1616--508 whose
  center is located  $10'$
west of the pulsar.
We find no preferential extension of the X-ray PWN toward the TeV source.
Therefore,
the {\sl Chandra} data do not provide conclusive
evidence for PSR J1617--5055 and  HESS J1616--508
  association.
We have also analyzed archival X-ray, radio, and IR data on the
HESS J1616--508 region and found traces of diffuse emission (resembling a shell in the radio) coinciding with the
  central part of HESS J1616--508. We speculate that the TeV source may be multiple, with most of the emission coming from an unknown SNR or
   a star-forming region,
 while some fraction of the TeV emission still may be attributed to the J1617 PWN.
\end{abstract}


\section{Introduction}
 High-resolution X-ray observations of young
($\tau\lesssim 10$ kyrs) rotation-powered pulsars usually
show a pointlike pulsar
embedded in
a pulsar wind nebula (PWN).
In most cases, the pulsar emission
is dominated by the nonthermal emission generated in the pulsar
magnetosphere. The X-ray spectrum of this emission usually  fits
power-law model with typical photon indices
$\Gamma\simeq1$--$2$. As a rule, the nonthermal  emission component  is
 strongly pulsed, showing one or more peaks per period.
Studying the  non-thermal
 emission constrains
the emission and particle acceleration mechanisms in the pulsar magnetosphere.
Young pulsars are also  interesting because they emit strong
magnetized relativistic winds which interact with the ambient
medium and form PWNe whose emission is interpreted  as
synchrotron radiation from the shocked pulsar wind (see Kaspi et
al.\ 2006; Gaensler $\&$ Slane 2006; Kargaltsev \& Pavlov 2008,
hereafter KP08, for reviews). Studying PWNe helps to understand
the structure and dynamics of relativistic pulsar winds, elucidate the
mechanisms of PWN formation, evolution and interaction with the
ambient medium, and establish the properties of the relativistic
plasma in PWNe.

 The young  ($\tau
\equiv P/2\dot{P} = 8.13$ kyr; $\dot{E}=1.6\times10^{37}$ ergs
s$^{-1}$) 69 ms pulsar J1617--5055 (hereafter J1617) was first
identified through its X-ray pulsations with {\sl ASCA} (Torii et al.\
1998)\footnote{The 69 ms pulsations from this region were first
detected with {\sl Ginga} (Aoki et al.\ 1992), but it was impossible to
determine the source of these pulsations because of the very poor
angular resolution of {\sl Ginga}.}
    with the  radio pulsations found shortly afterwards (Kaspi et al.\ 1998).
The {\sl ASCA} spectrum of the pulsar
was described by the absorbed  power-law (PL) model with the photon
index $\Gamma=1.4\pm0.2$
     and
     observed flux $F_X=(3.6\pm0.2)\times 10^{-12}$ ergs s$^{-1}$ cm$^{-2}$
in the 3$-$10 keV band (Torii et al.\ 2000).

Since the {\sl ASCA} detection, J1617 has been observed with other
X-ray satellites.
It was imaged
 $6'$ off-axis in {\sl Chandra} ACIS
observations of the RCW 103 supernova remnant (SNR)
 in 1999 September (Garmire et al.\ 1999)
and February 2000.  It was also observed
by {\sl XMM-Newton} in 2001 September (two pointings,
30 ks aimed at the pulsar,  and 20 ks centered at  RCW 103).
The J1617 properties inferred from these {\sl XMM-Newton} observations
   have  been briefly reported by
 Becker \& Aschenbach (2002; hereafter BA02). They found
 that the spectrum in the 0.5--10 keV band fits the
 absorbed PL model with $\Gamma=1.1$--1.4,
$n_H=(2.8$--$3.6)\times10^{22}$ cm$^{-2}$,
 and
   unabsorbed flux
$F_X^{\rm unabs}=(4.9$--$5.4)\times10^{-12}$ ergs$^{-1}$ cm$^{-2}$
  (the numbers correspond to the 90\% confidence range).
BA02 have also detected pulsations with
a single, asymmetric pulse and a pulsed fraction
of $\sim 50\%$ in the 2.5--15 keV band.
These observations confirmed the non-thermal nature of the pulsar
emission;
 however, they lacked the resolution needed to
separate the pulsar emission from the emission of a possible
compact PWN around this young, remote pulsar.
J1617 was also observed
serendipitously by {\sl XMM-Newton} in 2005 August ($\approx
90$ ks).

  J1617 has been  also  detected at higher energies with {\sl RXTE}
PCA/HEXTE (2$-$60 keV band; Torii et al.\ 2000; Kuiper 2007), and {\sl
Integral}
 IBIS/ISGRI
  (20--300 keV band; Landi et al.\ 2007).
    Landi et al.\ (2007) have produced a joined
  spectral fit
 to the {\sl XMM-Newton}, {\sl BeppoSAX}, and {\sl Integral} data and found
fitting parameters close to those obtained by BA02
from the {\sl XMM-Newton} data alone [$\Gamma=1.4\pm0.1$,
  $n_H=(3.9\pm0.3)\times10^{22}$ cm$^{-2}$, and
$F_X^{\rm unabs} =4.2\times10^{-12}$ ergs $^{-1}$ cm$^{-2}$ in
2--10 keV].
Fitting the {\sl RXTE} spectrum in the 2$-$30 keV band
with a power-law model, Kuiper (2007) found
$\Gamma=1.30\pm0.01$ for fixed $n_{\rm H}=3.2\times10^{22}$
  cm$^{-2}$.

\begin{deluxetable*}{lllll}
 \tablewidth{0pt}
\setlength{\tabcolsep}{0.1in}
 \tablecaption{Chandra Observations of J1617.}
\tablehead{\colhead{ObsID} & \colhead{123} & \colhead{970}  &
\colhead{2759} & \colhead{6684}}
\startdata Date & 1999 Sep 26  & 2000 Feb 8  &2002  March 03 & 2006 June 6 \\
Mode, chip region  & TE, Full Frame & TE, Full Frame & CC, Full Frame & TE, 1/4-subarray   \\
Telemetry format & VF & F & F & VF \\
 Chips activated
&I0-I3, S2,S3 & I2,I3,S2-S4 & I2,I3,S1-S4 & I3  \\
J1617 imaged on
& I1 & I3 & S2 & I3    \\
Off-axis angle & $6.1'$ & $6.1'$ & $6.3'$ & $40''$   \\
 Sci.\ Exp., ks
& 13.92 & 13.75 & 48.82 & 57.24 \\
\enddata
\tablecomments{ Except for ObsID 970, the scientific exposures
coincide with the live times given in the EVT2
 file headers (for the chip of
interest). In ObsID 970, removing the flares reduced this time  by
$\approx5$ ks. }
\end{deluxetable*}

Observations of J1617 and its surroundings have
become particularly interesting
after the
discovery of the
   extended ($\sim20'$ in diameter) TeV source HESS\,J1616--508
(hereafter HESS\,J1616) by Aharonian et al.\ (2006). Although its
center of gravity is offset by $\approx10'-11'$  from the pulsar,
  the
extent of the TeV emission still encompasses the pulsar location.
The HESS\,J1616 spectrum above 200 GeV  fits a power-law model with
  $\Gamma=2.35\pm0.06$ and 1--10 TeV flux $F_{\rm TeV}\approx
1.7\times10^{-11}$ ergs s$^{-1}$ cm$^{-2}$.
The central region of HESS\,J1616 was observed with {\sl Suzaku} XIS,
but no X-ray counterpart was found to a limiting flux of $3.1\times
10^{-13}$ ergs cm$^{-2}$ s$^{-1}$, so that the TeV source was
suggested to be ``the best example in the Galaxy of a dark particle
accelerator'' (Matsumoto et al.\ 2007). Landi et al.\ (2007) observed
the field of HESS\,J1616 with {\sl Integral} IBIS/ISGRI  and {\sl Swift}
XRT, and also analyzed archival {\sl XMM-Newton} and {\sl
BeppoSAX} data. These authors concluded that J1617 is the only
plausible counterpart to the TeV source.

    For many young pulsars,
radio or X-ray SNR
   associations have been found,
but no SNR association has been established for J1617 so far.
      Although J1617
    is just $\sim7'$
north of the center of the RCW 103 SNR, outside of the $5'$ SNR
radius,
the pulsar is almost certainly unrelated to the SNR.
RCW 103 hosts a central compact object (CCO; possibly
    a NS binary or a transient magnetar; e.g., Pavlov et al.\ 2002, 2004;
	de Luca et al.\ 2007), which is a more plausible candidate for the
compact remnant of the SN explosion. In addition, the RCW 103
spectrum shows a significantly smaller interstellar absorbtion than
J1617 (Gotthelf et al.\ 1997).
The relative positions and sizes of HESS\,J1616 and
 RCW~103
 suggest that the HESS source is also not associated with
 the SNR.

To search for a compact X-ray  PWN
around J1617 and separate the pulsar and PWN emission, we have carried
out a deep {\sl Chandra}
ACIS observation of J1617 in 2006 June, with the
pulsar placed near the optical axis. Here we report the
 results of this observation and our analysis
of archival multiwavelength data on J1617 and HESS\,J1616.
In \S2 we describe the observations, the images of the
 J1617 PWN and the HESS~J1616 field,
and the  spectral analysis  of the pulsar
and the PWN. We discuss the pulsar and PWN properties and the nature of
HESS\,J1616 in  \S3, and summarize our findings in \S4.

  \begin{figure}
 \centering
\includegraphics[width=3.2in,angle=0]{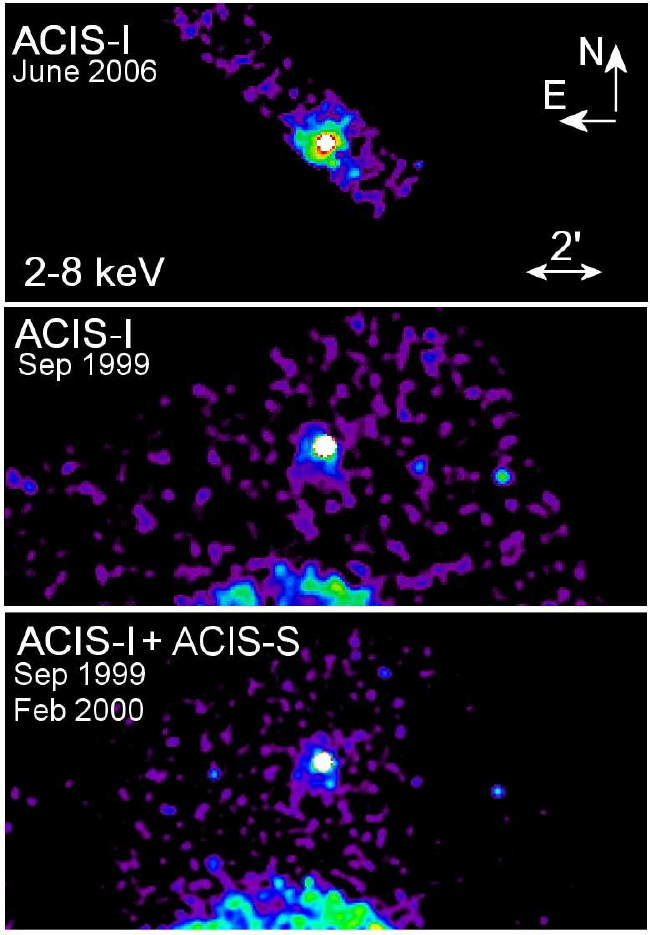}
\caption{{\sl Chandra}  images of the field around J1617 with
the observation dates listed in the images (see Table 1 for details).
The  {\em bottom}
panel shows the combined image from 1999 and 2000 observations.
All the images are binned to a pixel size of $4''$ and
then smoothed with the $r=12''$ gaussian kernel.
The contrast is chosen to emphasize the faint large-scale emission.
The extended shell, partly  seen in the {\em middle} and
{\em bottom} images,
 belongs to RCW 103.
 }
\end{figure}

\section{Observations and data analysis}

We observed J1617 with the {\sl Chandra} ACIS detector on 2006 June 6
(ObsID 6684) for 60 ks in Timed Exposure (TE) mode with
Very Faint (VF) telemtry format.
The pulsar was imaged on the ACIS-I3 chip near the optical axis.
The other ACIS chips were turned off.
To reduce the pile-up in the pulsar image,
we used 1/4 subarray ($\approx8'\times 2'$ field of view [FOV];
frame time 0.84104 s,
including 0.8 s exposure time and 0.04104 s ``dead time'').
No substantial
background changes were
 detected during the entire observation.
The scientific exposure time was 57.24 ks.

\begin{table}[]
\caption[]{Count statistics for the spectral extraction regions}
\vspace{-0.7cm}
\begin{center}
\begin{tabular}{cccccccc}
\tableline\tableline Region & $A$   &
 $N_{\rm tot}$  &
 $N_{\rm bg}$   &  $N_{\rm src}$ & S/N & ${\mathcal S}$ \\
\tableline
          inner PWN           &
         3.14 &  $1162\pm32$     &  $732\pm 75$   &  $430\pm82$ & 5.2  & $2.39\pm0.46$ \\
         outer PWN        &
       2610  &  $959\pm31$     & $445\pm21$    &   $514\pm37$ & 13.9  & $(3.44\pm0.25)\times10^{-3}$  \\
      \tableline
\end{tabular}
\end{center}
\vspace{-0.3cm} \tablecomments{Source ($N_{\rm src}$), total
($N_{\rm tot}$) and background ($N_{\rm bg}$) counts  are
extracted from the regions with an area $A$ (in arcsec$^{2}$) shown
in Figures 2 and
 4, in the 0.5--8 keV band. For {\em inner} PWN the
uncertainty of the background, $\simeq 10$\%, is mostly systematic
(due to the uncertainties involved  in the PSF simulation). The mean
surface brightness ${\mathcal S}$ is in units of
 counts ks$^{-1}$  arcsec$^{-2}$ in the 0.5--8 keV band.
} \vspace{0.2cm}
\end{table}

We will also use in this work two archival {\sl Chandra} ACIS imaging
observations
of 1999 September 26 (ObsID 123) and 2000 February 8 (970). As the
primary target of those observations was the RCW 103 SNR, the pulsar
was imaged $\approx6'$
off-axis, which resulted in a broadened point
spread function (PSF). The frame time was 3.24104 s in both
observations
(other details are given in Table 1.)
The second observation suffered from
 significant background flares (up to 6 times the quiescent level). Filtering
 out the flares reduced the
exposure by $\approx5$ ks.

In addition we analyzed the ACIS observation of
of 2002 March 3 (ObsID 2759). The data were
taken in Continuous Clocking (CC) mode, which allows one to
achieve time resolution of 2.85 ms at the expense of spatial
information in one dimension.

  The central part of HESS\,1616 was serendipitously imaged on
the ACIS S2 and S3 chips
during the {\sl Chandra} observations
 of the Kes 32 SNR that occurred on 2001 October 10
(ObsID 1960).  The
  data were taken in Faint Mode, and
the
scientific exposure time was 29 ks (Vink 2004).

Reduction of the {\sl Chandra} data
 was done with the {\sl Chandra} Interactive Analysis Observations (CIAO)
software (ver.\ 3.4, 4.0; CALDB ver.\ 3.3.0.1, 3.4.0)
 and FTOOLS (ver.\ 6.3).
    In our image analysis we have made use of MARX\footnote{MARX (Model of AXAF
Response to X-rays) is a suite of programs designed to enable the user
to simulate the on-orbit performance of the {\sl Chandra} satellite.
See \url{http://space.mit.edu/ASC/MARX/}} and {\sl Chandra} Ray Tracer
(ChaRT) software\footnote{The sofware is
 available at \url{http://cxc.harvard.edu/chart/}.}.
We also used  XSPEC (ver.\ 11.3.2) for the
spectral analysis.

J1617 was first  observed with the {\sl XMM-Newton} EPIC MOS1 and
MOS2 detectors
on 2001  September 9
(ObsID 0113050701)\footnote{J1617 was also serendipitously imaged
off-axis in the observation 0113050601 on the same date. We do not
 analyze this observation because a much longer {\sl XMM-Newton} observation of 2005 August
  is available, with the about the same pointing and roll angle.}.
  J1617 was also serendipitously observed with
the same detectors during the observation of
  RCW 103 on 2005  August 23 and 24 (ObsID 0302390101).
In both observations
 the two  MOS cameras were operated in Full Window mode
(2.6 s time resolution) with the Medium optical filter;
the dead-time corrected exposure times were
 $\approx28$  and  $\approx86$ ks.
 During the 2001 observation,
the EPIC PN camera was in Timing mode,
 while it was in Small Window mode, with J1617 outside of the FOV,
 during the 2005 observation.
For the {\sl XMM-Newton} data analysis,
 we used the Processing Pipeline Subsystem (PPS). Both
the 2001 and 2005 observations were strongly contaminated by
flares during which the background count rate exceeded the quiescent
 level by a factor of up to 16. Filtering the flares out
left  22.2 ks and 70.3 ks useful scientific
 exposure for the
first and second observations respectively.

  \begin{figure}
 \centering
\includegraphics[width=3.2in,angle=0]{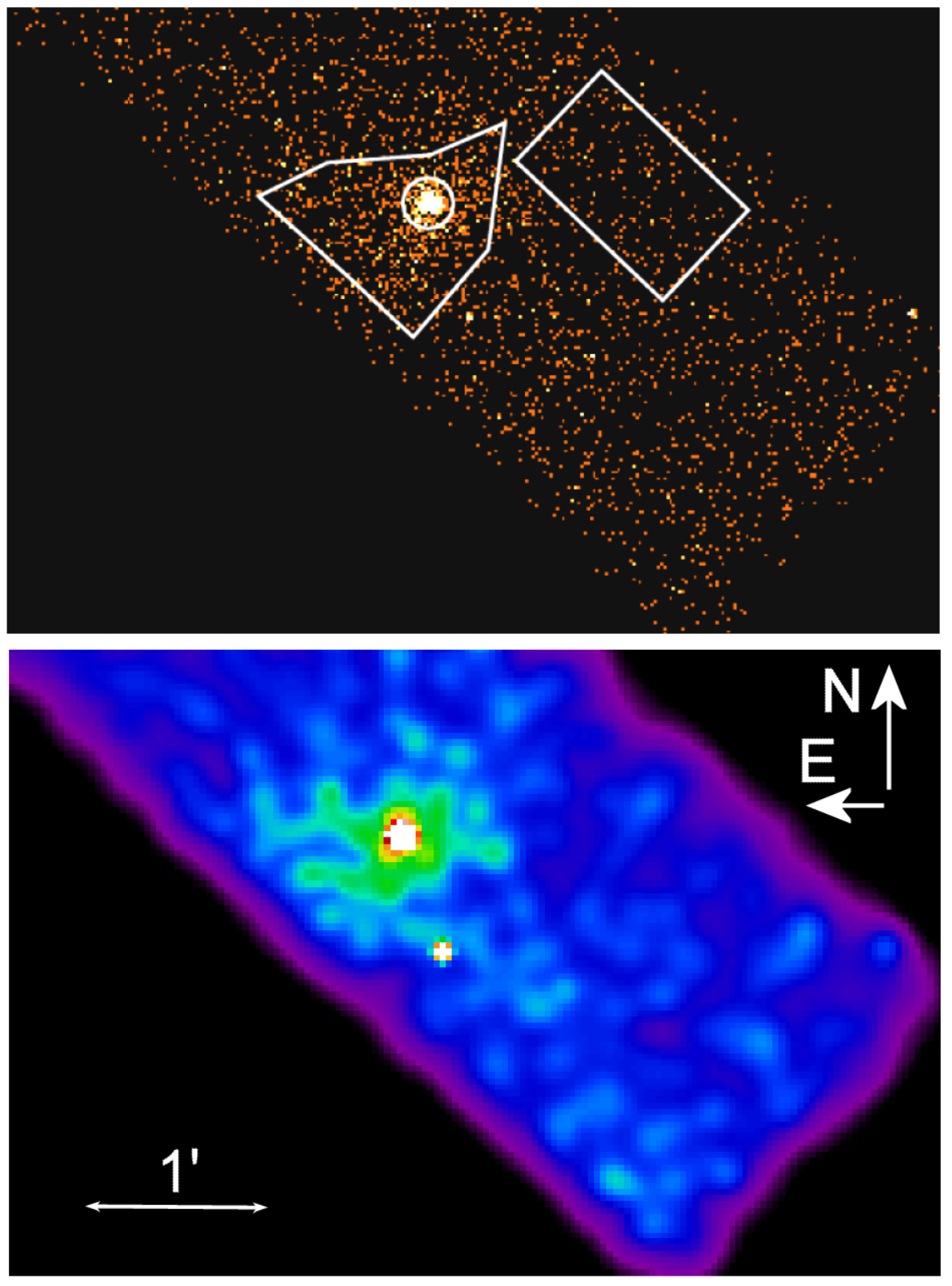}
\caption{ACIS-I3 (1/4 subarray) image of J1617 and its vicinity in the
2--8 keV band, binned by a factor of 2 (pixel size $0.98''$).
{\em Top}: Unsmoothed image with the regions used for extraction of the
extended PWN spectrum (within the polygon, excluding the circle;
see \S2.3.2 and Tables 2 and 3) and the
background (box).
{\em Bottom}: The same image adaptively smoothed with $\tt csmooth$.
}
  \label{acis_large}
\end{figure}

\subsection{PWN and pulsar images}

Figure 1
shows binned and smoothed large-scale
images of J1617 and its surroundings in the 2--8 keV band.
The upper panel shows the 1/4 subarray ACIS-I3 image  from our
observation of 2006 June.
In this image we see a relatively bright
pointlike source at the radio
pulsar position
 embedded in faint
 diffuse emission more extended toward south and southeast of the pulsar.
 The asymmetry is better seen in the ACIS images from earlier
observations (Fig.\ 1, {\em middle} and {\em bottom}),
 which have larger FOVs,
 with the pulsar
$\approx3'$ north of the RCW 103
 shell seen near the bottom in each of the two images.

 Figure 2 shows a close-up view of J1617 and its immediate vicinity
from our observation of 2006 June.
With the pulsar imaged on-axis,
this observation provides a much better resolution
than the off-axis ACIS observations of 1999 and 2000.
The pointlike
source, surrounded by diffuse emission, is
centered at R.A.$=16^{\rm h}17^{\rm m}29.35^{\rm s}$, decl.$=-50^{\circ}55'
12.8''$.
The position uncertainty is dominated by the uncertainty of
the {\sl Chandra} absolute astrometry, $0.65''$ at the 90\% confidence
level for sources within $2'$ from the optical axis on the ACIS-I3
chip\footnote{See \S5.4 and Fig.\ 5.1 of the {\sl Chandra} Proposers'
Observatory Guide v.\ 10 at \url{http://asc.harvard.edu/proposer/POG}.}.
As this position
differs by only $0.62''$
from the radio position of
J1617 reported by Kaspi et al.\ (1999), the
pointlike source must be the X-ray counterpart of the radio pulsar.
The  amorphous diffuse  emission around J1617 is
 distinguishable from the background up to $\approx 1'$ from the pulsar.
The background-subtracted
surface brightness of this extended PWN is
 $\simeq0.002$--$0.01$ counts arcsec$^{-2}$
ks$^{-1}$,
being somewhat brighter
 south and southwest of the pulsar.

\begin{figure}
 \centering
\includegraphics[width=3.2in,angle=0]{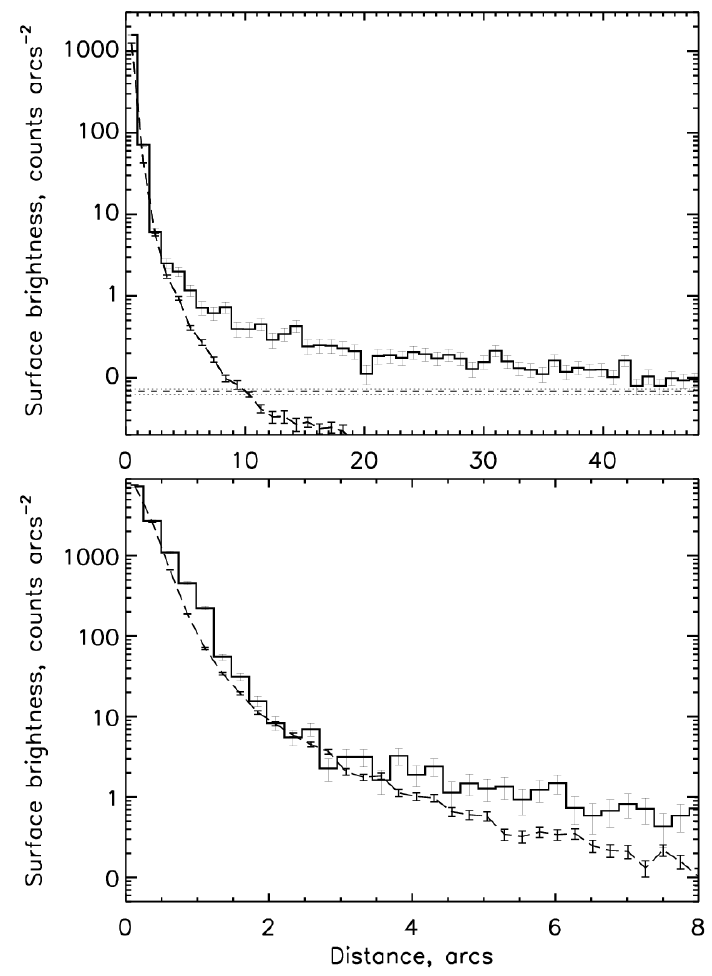}
\caption{Radial profiles of the observed emission around the pulsar position
and the simulated PSF in the 2--8 keV band.
 The histogram in the {\em top} panel (bin size $1''$) demonstrates the
large-scale PWN emission.  The histogram in the {\em bottom} panel
(bin size $0.25''$) shows the compact inner PWN contribution,
including the excess at
$\approx 0.5''$--$1.5''$ from the pulsar (see also \S2.1 and Fig.\ 4).
}
\end{figure}

The radial profile of the detected emission, centered on the pulsar position,
is shown in Figure 3,
  together with the simulated PSF\footnote{ To simulate the PSF with ChaRT
and MARX, we followed the steps
  outlined in \url{http://cxc.harvard.edu/chart/threads/index.html}. We have also tried several values of the Dither Blur parameter
  (see \url{http://cxc.harvard.edu/chart/caveats.html}
 and Misanovic et al.\ 2007 for discussion) and adopted the value of $0.1''$
that provides the best
  match between the simulated PSF and the real data within $r\lesssim3''$ from the pulsar. }
 and background level.
We see that the extended, faint PWN prevails over the pulsar emission
at distances $\gtrsim5''$ from the pulsar.

Given the large distance to the pulsar and the faintness of
the extended PWN,
one could expect a more compact PWN component, which would not
be immediately seen in pipeline-processed data.
 To search for
such a component,
we  produced images of the immediate pulsar vicinity
 at subpixel resolution (see Fig.\ 4).
We first  removed the
pipeline pixel randomization and applied  the sub-pixel resolution
tool to split-pixel events in the image
 (Tsunemi et al.\ 2001; Mori et al.\ 2001).
We then simulated the PSF image
in the
2--8 keV
 energy band
  at the same position on the detector as in the
 real observation. We binned the
observed image and the
simulated PSF by a factor of 0.2
 compared to the original ACIS pixel size.
  Finally, we used the CIAO ver.\ 4.0  implementation
(provided as an {\tt arestore} script)
 of the Lucy-Richardson deconvolution algorithm (Lucy 1974).
We ran {\tt arestore} with several settings for
 different numbers of iterations, between 30 and 200,
and noticed that the iterations converge
(i.e., no new
structure appear)
once the number of iterations
 approaches 100.

Examples of such deconvolved images are shown
 in Figure 4  for 30, 100, and 200 iterations
(panels {\em b, c}, and {\em d}, respectively).
When the number of iterations approaches 30,
 the deconvolved image
starts to reveal an extended structure ($\approx2''$ in
  linear extent) elongated in the south-north direction,
 with the point
   source being in the middle of this structure.
With increasing the number of iterations,
the elongated structure becomes
   one-sided (i.e., it appears only to the south of the point source),
and an  arclike
   feature, roughly perpendicular to the southern elongation,
 appears at about $1''$ north of the pulsar.
  Increasing the number of iterations beyond $\approx 100$
does not
change the structure of the compact extended emission.
The exposed  features
of the deconvloved images could be interpreted as a jet
(south of the
pulsar) and an arc (perhaps part of a ring or a torus).
Note that similar
  features are often seen in {\sl Chandra} images of PWNe around
young pulsars (see e.g., KP08). The radial profile shown
  in Figure 3 ({\em bottom}) demonstrates the significance of the count
excess at
$r\approx 0.5''$--$1.5''$
 from the pulsar.

For comparison, we also show
 a deconvolved
ACIS image without removing
the pipeline pixel randomization (Fig.\ 4{\em e}).
Although the deconvolution of this image does not
recover
as much structure as the
images with the pixel randomization removed, it
does  show some of the features that are seen in the
deconvolved non-randomized images.

 To test
reliability of the  deconvolution
procedure, we
 applied
it to the ACIS image of the CCO in the Cas A SNR
(ObsID 6690),
a radio-silent NS showing no radio pulsar
activity, for which no PWN is expected (e.g., Pavlov et al.\ 2004).
  The ACIS image of the CCO has $\approx 7200$ counts within the
$0.98''$ radius
around the source, comparable to J1617 in our ACIS-I image.
Unlike  J1617, the deconvolved image of
the Cas A CCO preserves the pointlike
  appearance with no extended structure (see Fig.\ 4{\em f}).
Although, as any deconvolution  algorithm, {\tt arestore} could introduce
   artifacts in the images,  the test example of
the Cas A CCO
provides
   credibility to the features seen in the deconvolved images of J1617.

    \begin{figure*}
 \centering
\includegraphics[width=7.0in,angle=0]{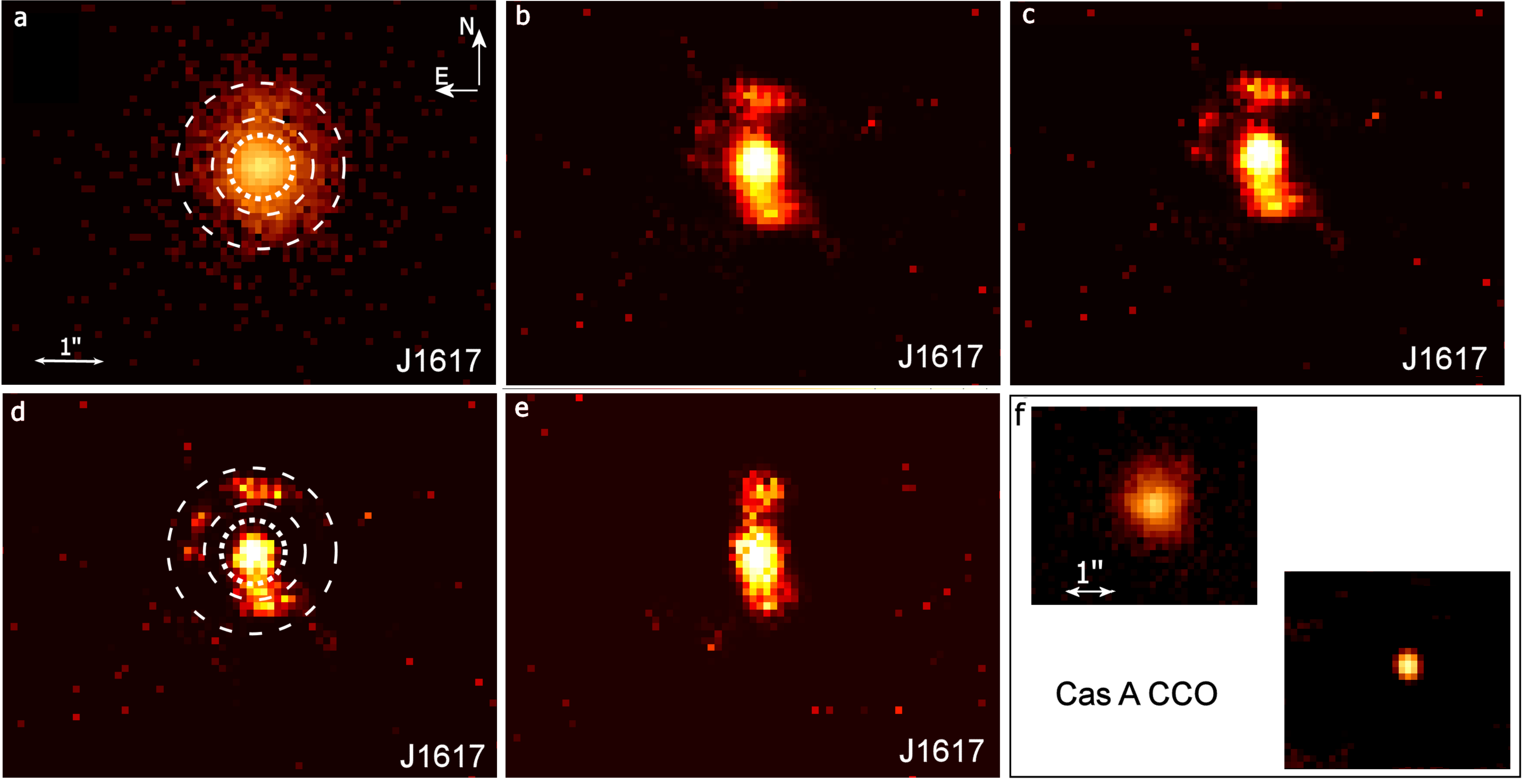}
\caption{
{\em a:} Original ACIS-I3 image of pulsar and its immediate
vicinty (pixel size $0.1''$), with
pipeline pixel randomization removed
  (circle radii are $0.5''$, $0.74''$ and $1.25''$).
{\em b}, {\em c}, and {\em d}:
 Deconvolved images obtained with {\tt arestore} from the image in panel
{\em a} after 30, 100, and 200 iterations, respectively.
{\em e:} Same as {\em c} but without removing
 the pipeline pixel randomization
in the original image.
{\em f:} Original (top left; pixel size $0.1''$, pipeline
randomization removed)
 and deconvolved (bottom right; {\tt arestore} with 100 iterations)
images of the Cas A CCO. See text for  details.}
  \label{deconv}
\end{figure*}

  \subsection{The field of HESS J1616--508.}

 Since J1617 is considered as a plausible counterpart
for the extended TeV source HESS\,J1616
(Landi et al.\ 2007),
 we have searched
 for additional
data in the {\sl Chandra} and {\sl XMM-Newton}
archives that would provide at least partial coverage
 of  the  HESS\,J1616 extent
and allow us to look for other potential X-ray counterparts of
the TeV source and examine its connection to J1617.
In the {\sl Chandra} archive we  found an
  ACIS observation of
the Kes 32 SNR (Vink 2004).
During that observation,
the S2 and S3
   chips
   imaged the
   central part of  HESS\,J1616 (see Fig.\ 5{\em a}).
The brightest source
(marked X in Fig.\ 5{\em a})
  is located on the S2 chip around
R.A.$\approx16^{\rm h}16^{\rm m}10^{\rm s}$,
decl.$\approx-50^{\circ}54' 30''$. Because of the small number of
counts ($140\pm 20$ counts within an $r=40''$ circle in 0.5--8 keV;
$S/N\approx7$) and the broad PSF (${\rm FWHM}\approx 15''$ at
the off-axis angle $\theta\approx16'$),
 it is difficult to determine whether the source is extended or multiple;
 however,
 with the linear extent of $\simeq1'$, it does not appear to
be point-like.
Assuming an absorbed PL spectrum with $\Gamma=1.5$ and
$n_{\rm H,22} \equiv n_{\rm H}/10^{22}\, {\rm cm}^{-2}=3.45$ (as
found from the fit to the pulsar spectrum; see \S2.3.1),
 the unabsorbed 0.5--8 keV flux of source X can be estimated as
 $F_{\rm X}\sim (1.7$--$2.2) \times 10^{-13}$ ergs cm$^{-2}$ s$^{-1}$.

We see some hints of
  extended emission southwest
 of source X;
 however, the low count rate,
$\sim 1.4$ counts  ks$^{-1}$ arcmin$^{-2}$ in 0.5--8 keV
 (corresponding to the unabsorbed surface brightness of $\sim6\times
10^{-14}$ ergs cm$^{-2}$ s$^{-1}$ arcmin$^{-2}$ for
the PL model
with $\Gamma=1.5$ and $n_{\rm H,22}=3.45$)
 precludes
 any detailed analysis of this
  diffuse emission.

  Source X falls near the edge the  EPIC FOV in the 2001 observation
(Fig.\ 5{\em b}).
   Although some emission
at the source X position is discernible,
 it is not possible to
determine
 whether the source is extended or pointlike.
Its brightest
 part in the EPIC image is
shifted by about $30''$ southwest from the {\sl Chandra} position
  (the difference could result from a statistical noise
 and background fluctuations).

 Source X was outside the EPIC FOV in the 2005 {\sl XMM-Newton} observation.
Because of the low surface brightness and large  XIS PSF, the source is
not seen  in the {\sl Suzaku} data.
 Deeper
ACIS or EPIC observations, with the center of the TeV source
imaged close to the optical axis, are required to understand the nature
  of the detected extended X-ray emission.

 In addition to analyzing the X-ray images,
 we examined the available IR and
radio data on the region of interest. The 843 MHz  image from the
  Sydney University Molonglo
 Sky Survey (SUMSS) shows some diffuse emission surrounding the
very central part of HESS\,J1616
  (within the innermost TeV contour shown in Fig.\ 5{\em a,f}).
 The radio emission
looks like a patchy,
 elongated shell, possibly an unknown SNR (alternatively,
 it could be
several faint point sources accidently aligned in a shell-like structure).
The {\sl Spitzer}
IRAC images of the same region (Fig.\ 5{\em e,f}) also
 reveal  diffuse emission partly coincident with the radio emission.
In addition, there are
several compact
sources seen
in the radio and, especially, IR images
  within the central part of the TeV source.
The brightest source, marked  A in Fig.\ 5,
  is located near the TeV source center.
At 8 $\mu$m, source A is resolved into a complex shell
  with an additional lobe to the west and a number
of compact point sources suggesting a star-forming region.
Source X appears  to have IR counterpart(s)
  resolved into several compact diffuse sources at $8$ $\mu$m (Fig.\ 5{\em e}), possibly also star-forming
  regions.
Deeper radio observations with higher
spatial resolution would be most helpful to understand the nature
of these sources.

 \begin{figure*}
 \centering
\includegraphics[width=6.0in,angle=0]{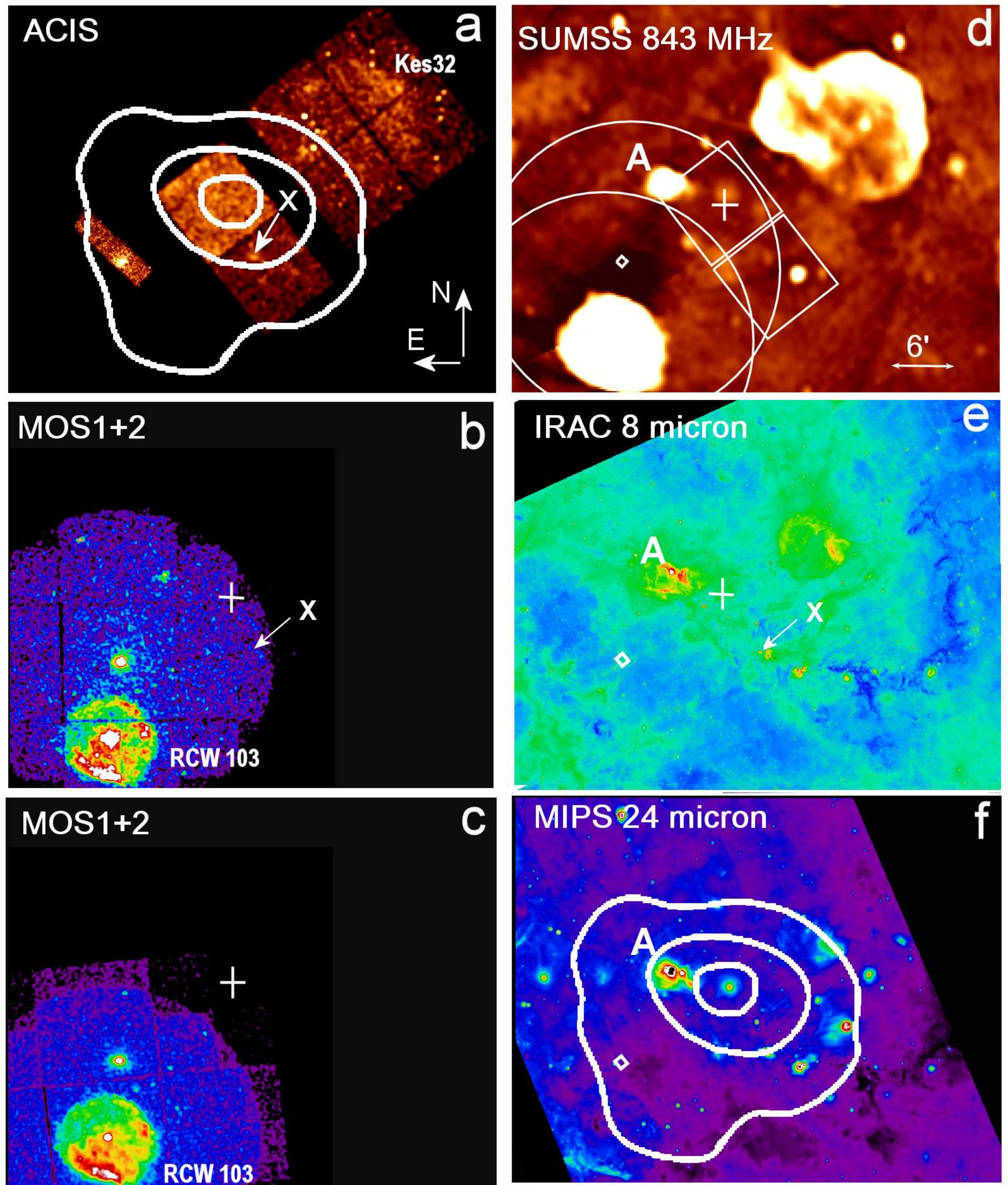}
\caption{ X-ray, radio and IR images of the J1617 and HESS\,J1616 field.
{\em a}: Combined 0.5--8 keV image from two ACIS observations
(2006 June and 2001 October) with
TeV contours (from Aharonian et al.\ 2006) overplotted.
{\em b} and {\em c}:
MOS1+2 images in the 0.5--8 keV band from the {\sl XMM-Newton }
observations of 2001 September
 and 2005 August, respectively.
 The cross shows
the  center of HESS\,J1616.
{\em d}:
843 MHz SUMSS image.
The rectangles show the region covered by the ACIS S2 and S3 chips
in the {\sl Chandra} observation of 2001 October, while the
circles show the MOS1+2 FOV in the two {\sl XMM-Newton} observations.
 The position of J1617 is shown by the diamond.
{\em e} and {\em f}: {\sl Spitzer} images of the field.
}
\label{multiwave}
\end{figure*}

\subsection{Pulsar timing}

 The data from the CC-mode observation of
the RCW 103 CCO also allow one  to analyze the pulsations
of the J1617 pulsar
  since the pulsar falls within the FOV.
The pulsar is clearly seen in the one-dimensional image shown in Figure 6;
 however, the large off-axis
   angle ($\approx6'$) significantly broadens the PSF,
 making it impossible to separate the inner PWN  and the pulsar.

 \begin{figure}
 \centering
\includegraphics[width=2.6in,angle=90]{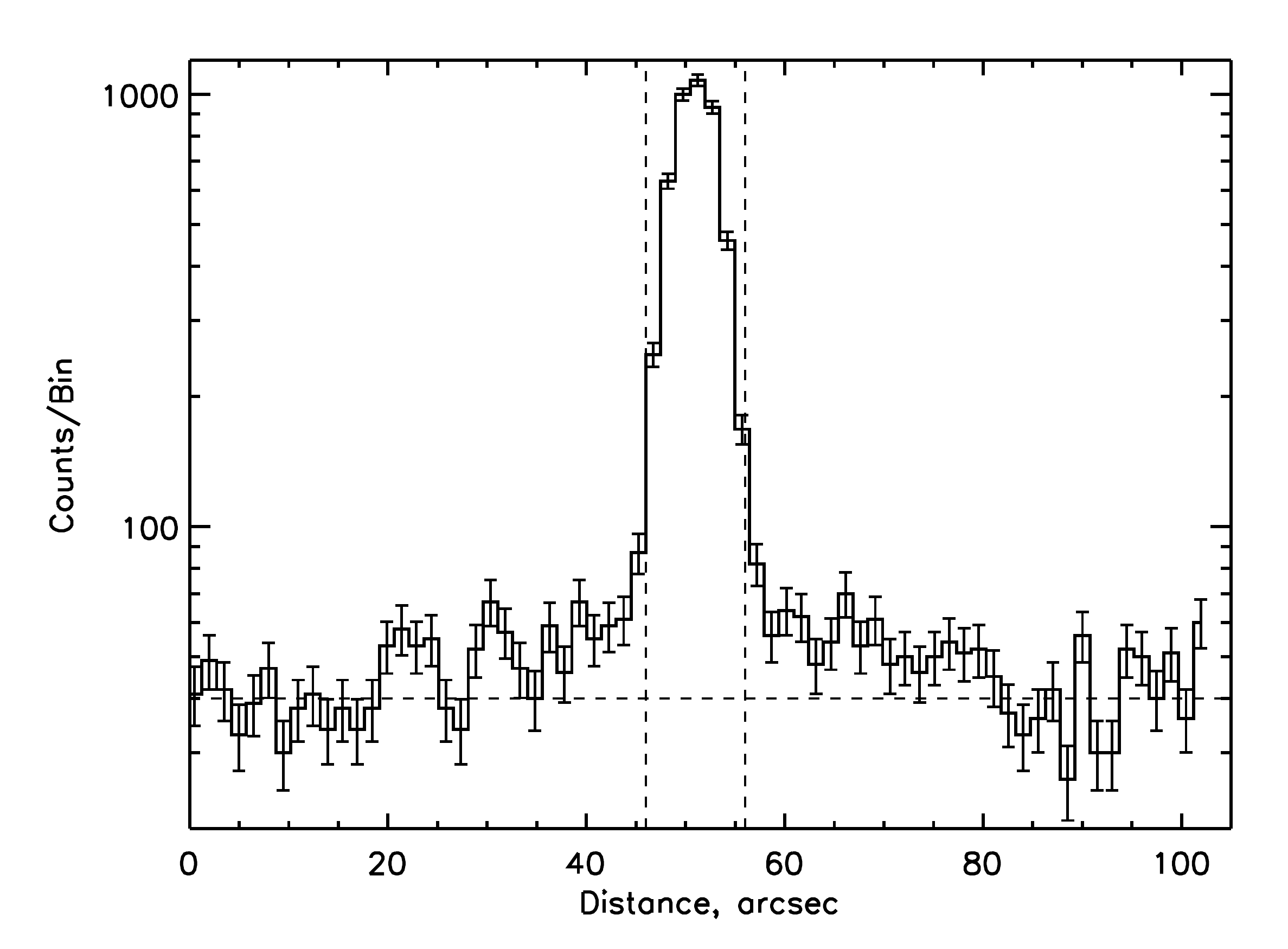}
\includegraphics[width=2.6in,angle=90]{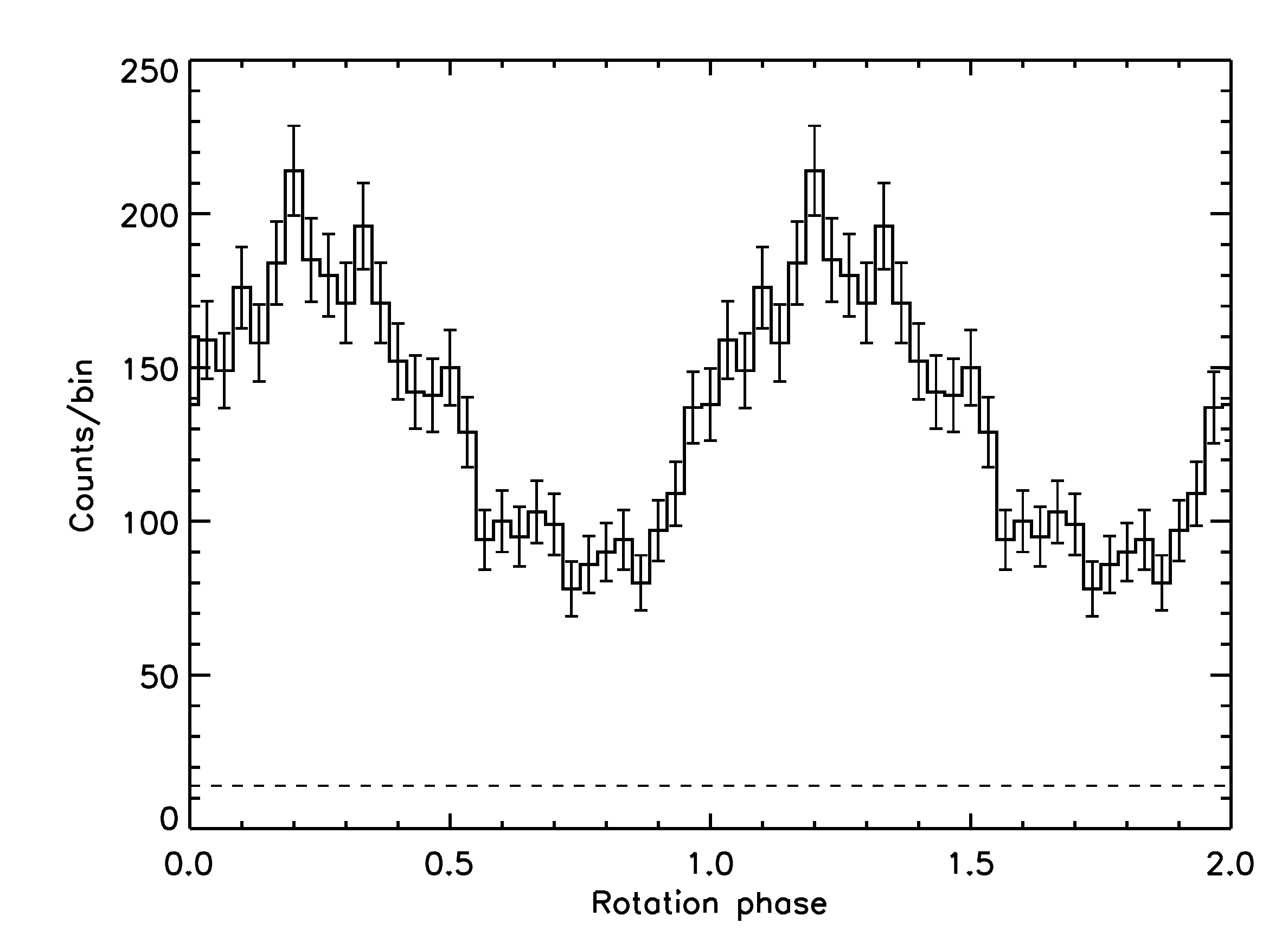}
\caption{ Results from the CC-mode observation. {\em Top}:
One-dimensional profile showing J1617 and the outer PWN above the
horizontal dashed line that corresponds to the background level.
The two vertical lines show the aperture from which the spectrum
and the light curve were extracted.
{\em Bottom}: The light curve with 30
phase bins.
The dashed line shows the background level (including the contribution
of the outer PWN).}
\end{figure}

For the timing analysis, we extracted 4057 photons in the 2--8 keV
band from
the $10''$-wide segment centered on J1617 (shown in Fig.\ 6, {\em top};
$\approx 90\%$
 of these counts are expected to come from the
unresolved source, i.e., the pulsar and the compact inner PWN).
  The photon
arrival times have been transformed to the solar system barycenter
using the {\tt axBary} tool of CIAO~3.4.
 We have searched for the pulsed signal near the expected
radio pulsation frequency and
 found $Z_{\rm 1,max}^{2}=301$
(see Buccheri et al.\ 1983 for details on $Z_n^2$ statistics)
at $\nu=14.414488\, {\rm Hz}\pm 1\,\mu{\rm Hz}$,
 consistent with the
  frequency expected from the radio ephemeris at the epoch of
the {\sl Chandra} observation.
   The light curve folded at this frequency reveals a single pulse with
a flat
   minimum  (see
  Fig.\ 6, {\em bottom}).
The observed 2--8 keV pulsed fraction  is
$40\%\pm 4\%$
Correcting it for the background contribution (which includes the outer PWN),
we obtain $44\% \pm 4\%$.
    The pulse profiles extracted in narrower energy bands (2--4 and 4--8 keV)
are similar
     to the 2$-$8 keV pulse profile, both in shape and pulsed fraction.

\subsection{Spectral analysis}

\subsubsection{Pulsar}

 For our 1/4 subarray observation (ObsID 6684),  we extracted the
pulsar's spectrum (using the CIAO's {\em psextract} task)
from a small circular aperture of
$r=0.5''$  to minimize the
contamination
from the compact PWN (see Fig.\ 3, {\em bottom}, and also Fig.\
4{\em a,d}). No background subtraction was attempted because
 the
PWN contribution is expected to be negligible in this small aperture
(Fig.\ 3, {\em bottom}).
 The 2853 counts extracted  from the $0.5''$
radius aperture ($52\%$ encircled energy fraction) in the 1--8 keV range
(there are virtually no counts below 1 keV) correspond to the
aperture-corrected  source
       count rate of
$96\pm 2$ counts ks$^{-1}$.
 The corresponding absorbed flux is
 $F_{\rm psr}\simeq2.15\times10^{-12}$ ergs cm$^{-2}$ s$^{-1}$.
With the (exposure) frame time of 0.8 s,
the expected pile-up fraction is very small ($\approx3\%$ according
to PIMMS\footnote{http://cxc.harvard.edu/toolkit/pimms.jsp}).

\begin{figure}
 \centering
\includegraphics[width=2.3in,angle=-90]{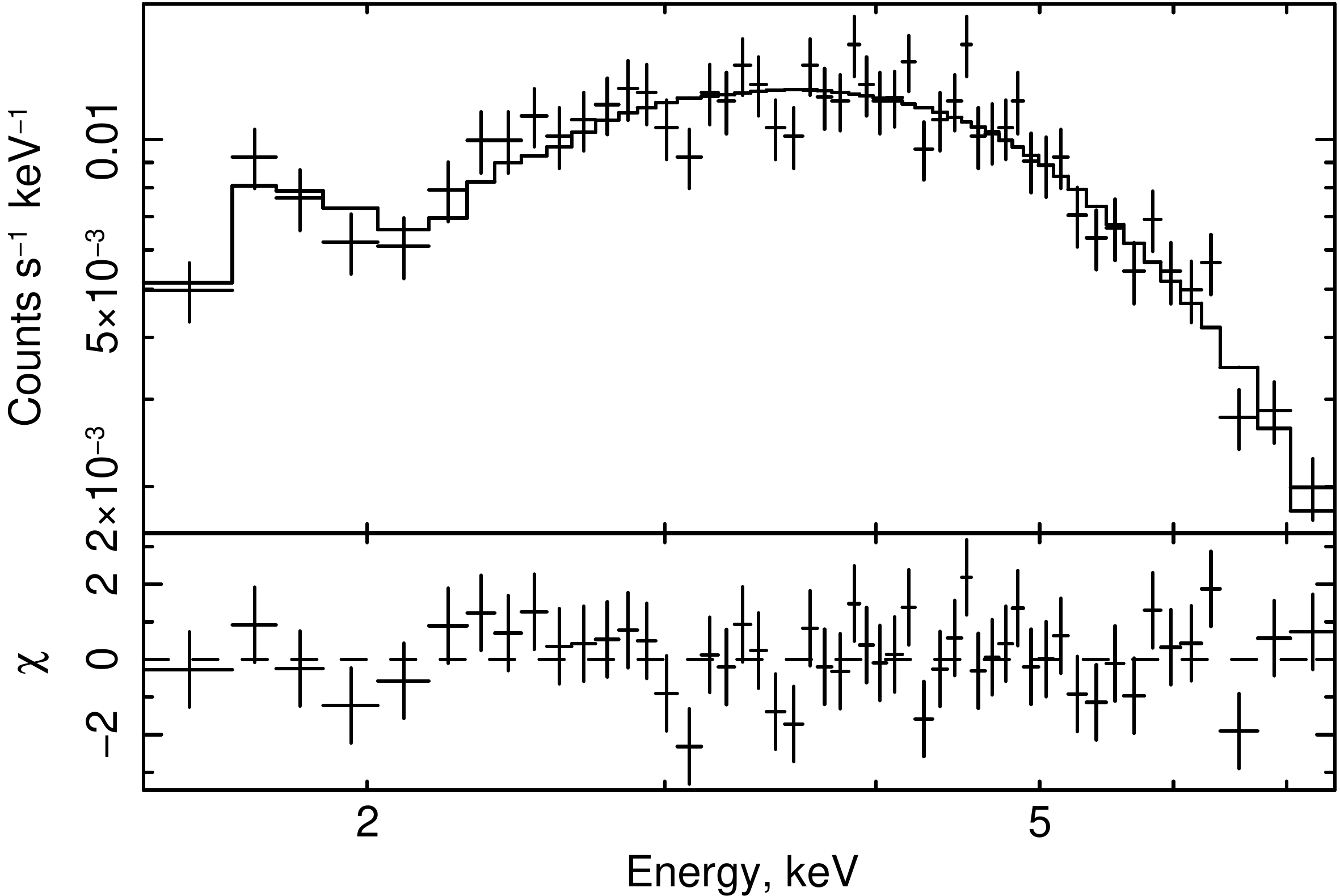}
\caption{Spectrum of the pulsar in observation 6684
 fitted with the PL model (see Table 3 for the fitting parameters).
}
\end{figure}

\begin{figure}
\includegraphics[width=3.2in,angle=0]{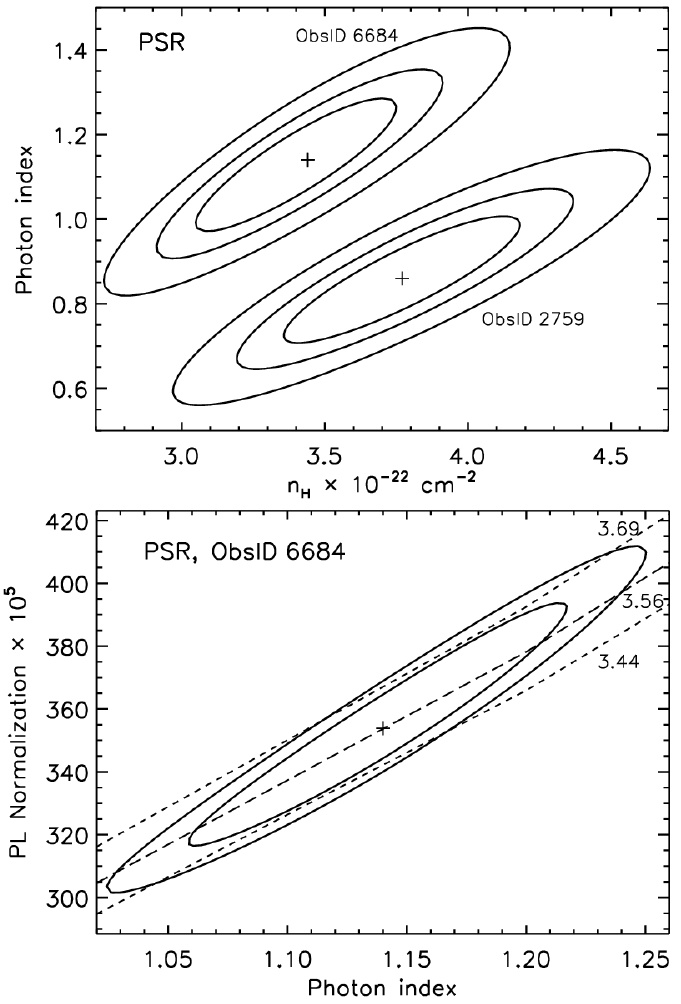}
\caption{  {\em Top}: Confidence contours (68\%, 90\%, and 99\%) in the $n_{\rm
H}$--$\Gamma$ plane for the PL fit to the pulsar spectrum from
two ACIS observations (see Table 1). The contours
 are obtained with the PL normalization fitted at each point
of the grid.
{\em Bottom}: Confidence contours (68\% and 90\%) for the PL fit to the
 pulsar spectrum
for fixed $n_{\rm H,22}=3.45$.
 The PL normalization
is in units of $10^{-5}$ photons cm$^{-2}$ s$^{-1}$ keV$^{-1}$ at 1
keV. The dashed curves are the loci
 of constant unabsorbed flux in the 0.5--8 keV band;
the flux values near the curves are in units of 10$^{-12}$ ergs
cm$^{-2}$ s$^{-1}$.
}
\end{figure}

We binned the spectrum with minimum of 50 counts per  bin and fit
the absorbed PL model in the 1--8 keV range with
all the fitting parameters allowed to vary.
The quality of the fit is excellent ($\chi_\nu^{2} = 1.04$
 for 49 degrees of freedom [dof]; see Fig.\ 7).
The hydrogen column density,
$n_{\rm H,22}=3.45\pm0.14$ (see Fig.\ 8; the errors here and below
are at the 68\% confidence level for a single  interesting parameter)
is a factor of 1.5
larger than
  the average Galactic
HI column density in the
  direction of the pulsar ($l=332.50^\circ$, $b=-0.28^\circ$),
$n_{\rm HI,22}=2.26$
(Dickey \& Lockman 1990).
This is not surprising since the $n_{\rm H}$ deduced from an X-ray
spectrum under the
assumption of standard element abundances generally exceeds the
$n_{\rm HI}$ measured from 21 cm observations by a factor of 1.5--3
(e.g., Baumgartner \& Mushotzky 2005).
   The same PL  fit gives the
photon index  $\Gamma_{\rm psr} = 1.14\pm0.06$ and
the unabsorbed flux $F_{\rm psr}^{\rm unabs}=
(3.6\pm0.1)\times10^{-12}$ ergs cm$^{-2}$ s$^{-1}$ in
the 0.5--8.0 keV band.
It corresponds to the isotropic luminosity
      $L_{\rm psr}\approx
1.8\times10^{34}d_{6.5}^2$ ergs s$^{-1}$
($=1.1\times10^{-3}d_{6.5}^2 \dot{E}$), where $d_{6.5}=d/6.5\,{\rm
kpc}$ is the distance to the pulsar estimated from
   the dispersion measure ${\rm DM}=467$ cm$^{-3}$ pc
and the Galactic electron density distribution model by Taylor \&
Cordes (1993)\footnote{The more recent
model by Cordes $\&$ Lazio (2002) gives a slightly larger distance of
6.8 kpc.
    We caution, however,  that distances based on such models
are not very certain. For instance,
HI absorption measurements give $\sim 20\%$--30\% smaller
distances than the Taylor \& Cordes model for two pulsars within
20$^\circ$ of J1617 (Kaspi et al.\ 1998).}.

\begin{table}[]
\caption[]{PL fits to the pulsar and PWN spectra. } \vspace{-0.5cm}
\begin{center}
\begin{tabular}{cccccccc}
\tableline\tableline
Region & $n_{\rm H,22}$\tablenotemark{a}   &
$\mathcal{N}_{-5}$\tablenotemark{b}  &
$\Gamma$ & $\chi^2$/dof   & $L_{\rm X,33}$\tablenotemark{c} & $I_{-16}$\tablenotemark{d}  &   \\
\tableline
   Pulsar  &  $ 3.45\pm0.14$  &  $35.4^{+7.3}_{-4.8}$  &  $1.14\pm0.06$   &  $51.0/49$  & $17.92\pm0.07$     &   ...   \\
Inner PWN   &  $[3.45]$  &
$4.2^{-1.0}_{+1.4}$   &   $1.47\pm0.24$   &  $17.5/19$  & $1.49^{+0.28}_{-0.23} $   & $940\pm160$         \\
 Outer PWN   &  $[3.45]$  &
$5.8_{-1.4}^{+1.7}$    &   $1.65\pm0.20$  &  $12.9/15$  & $1.71\pm 0.25 $   & $1.3\pm0.2$   \\
\tableline
\end{tabular}
\end{center}
\tablecomments{
 The uncertainties are given at the 68\%
confidence level for one interesting parameter. Systematic
uncertainties are not included. The normalization and luminosity of
the pulsar are corrected for the finite extraction aperture.}
\tablenotetext{a}{ $n_{\rm H,22}\equiv n_{\rm H}/10^{22}$
cm$^{-2}$ was fixed in the PWN fits. } \tablenotetext{b}{Spectral flux
in units of $10^{-5}$ photons cm$^{-2}$ s$^{-1}$ keV$^{-1}$ at 1
keV.} \tablenotetext{c}{ Isotropic luminosity  in units of $10^{33}$
ergs s$^{-1}$ in the 0.5--8 keV band, for $d=6.5$ kpc. }
\tablenotetext{d}{Mean unabsorbed intensity  in units of $10^{-16}$
ergs s$^{-1}$ cm$^{-2}$ arcsec$^{-2}$ in the 0.5--8 keV band.}
\end{table}

 We have also extracted the pulsar spectrum from
the ObsID 2957 data taken in the CC mode,
using
the same region ($10''$-wide segment) as for the light
curve (see \S2.3), which contains 4522 counts in the 1--8 keV band.
  Note that in this case
 the extraction region is much larger than that used for ObsID 6684
because
 the broad off-axis PSF. The background was taken from a nearby
($1.4'$--$2.4'$ from the pulsar)
region located on the same chip (S2).
  The background contribution is $\approx8\%$ in the 1-8 keV band.
  The absorbed PL fit gives $\Gamma\simeq0.85$ and $n_{\rm H,22}\simeq3.8$
(see Fig.\ 8 for the statistical uncertainties). Although
  the difference between the
$\Gamma$ values from ObsID 6684 and ObsID 2957 appears to be
   statistically significant, the CC-mode fit is likely
less reliable because of larger unaccounted systematic
   uncertainties
(the CTI, gain, and ACIS filter contamination corrections are
less accurate in the CC mode);
therefore,
below
we will use the more reliable
spectral parameters inferred from the ObsID 6684 data.

\subsubsection{PWN}

 We used the ObsID 6684 data to extract the PWN spectra
 from two regions.
The brightest
 {\em inner}
 component is extracted from
 the $0.75''<r<1.25''$ annulus with the area of 3.14 arcsec$^{2}$
(see Fig.\ 4).  In this case we used the pulsar spectrum as the background spectrum, after
 scaling the normalization according to the simulated PSF
shown in Figure 3 ({\em bottom}).
  The fainter {\em outer} component
  is  extracted  from a polygon region (see Fig.\ 2)
with an area of  2600 arcsec$^{2}$
(excluding the $r = 9''$ circle
around the pulsar).
  The background was taken from the $40'' \times 65''$ box
shown in Figure 2 ({\em top}).

The annulus region of the
 inner PWN component has 1162 counts (see Table 2 for details), with $\approx63$\%
of them  coming from the pulsar (treated as a background).
The corresponding absorbed flux is
$F_{\rm inner}=
(1.6\pm0.3)\times10^{-13}$
     ergs cm$^{-2}$ s$^{-1}$ in the 0.5--8 keV band.
      The outer PWN region contains 959 counts
(54\% coming from the source);
its absorbed flux is $F_{\rm outer}=(1.7\pm0.1)\times10^{-13}$ ergs
cm$^{-2}$ s$^{-1}$ in the same band.

\begin{figure}
 \centering
\includegraphics[width=2.7in,angle=90]{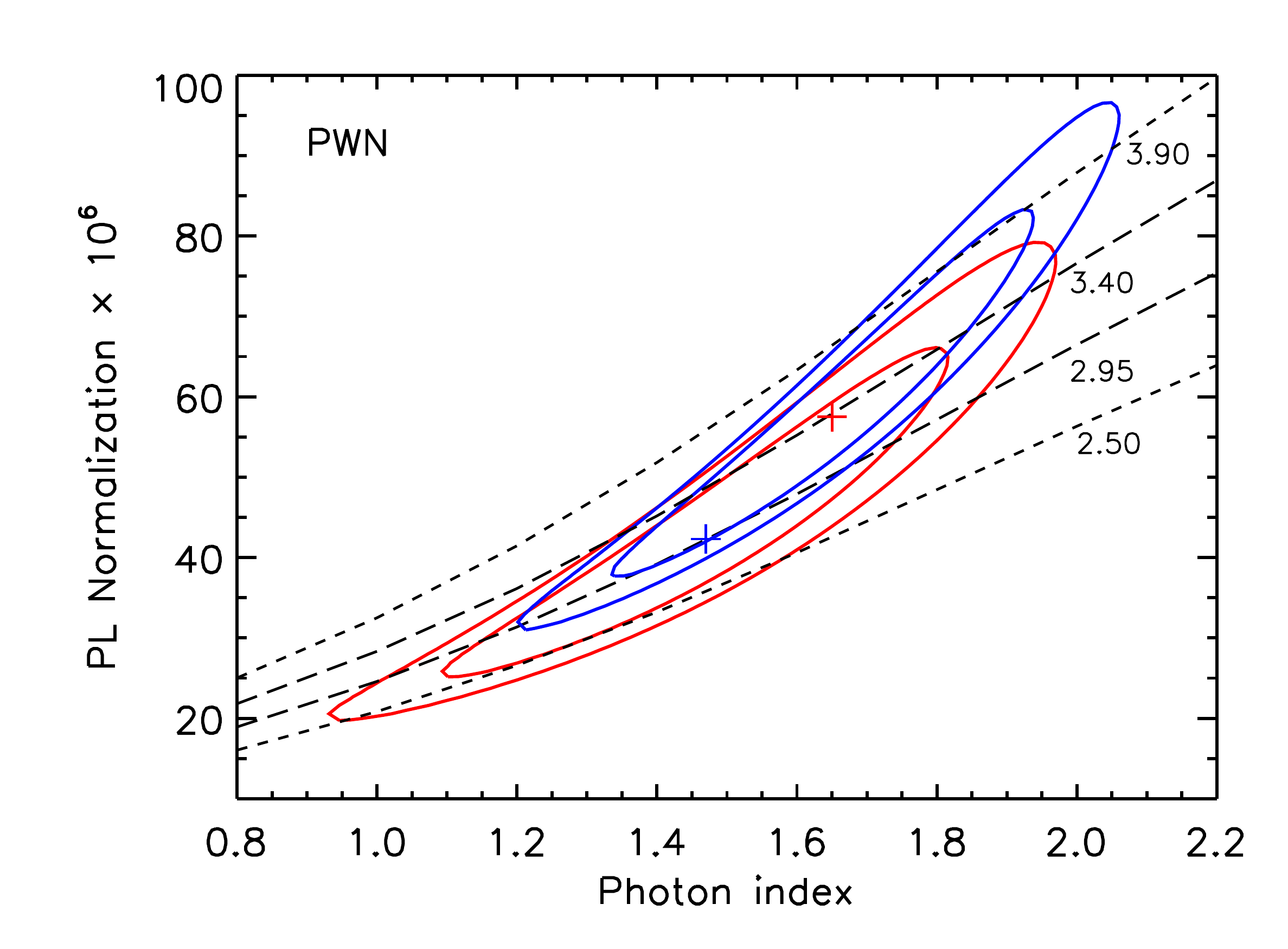}
\caption{  Confidence contours (68\% and 90\%) for the PL fit to the
 inner and outer PWN regions
(red and blue contours, respectively;
see text for the region definitions)
for fixed
 $n_{\rm H,22}=3.45$.
 The PL normalization
is in units of $10^{-6}$ photons cm$^{-2}$ s$^{-1}$ keV$^{-1}$ at 1
keV. The dashed curves are the loci
 of constant unabsorbed flux in the 0.5--8 keV band;
the flux values near the curves are in units of 10$^{-13}$ ergs
cm$^{-2}$ s$^{-1}$.
}
\end{figure}

 We
 binned the  spectra extracted from the inner
 and outer regions
with a minimum of 50 counts per spectral bin.
We
then fit the spectra
  with the absorbed PL model  with the hydrogen column density fixed at
the  best-fit value obtained from the PL fit to the pulsar spectrum,
$n_{\rm H,22}=3.45$.
 For the outer PWN component,
 we
restrict the energy range to 2--8 keV
because of the large contribution
  of the non-uniform background\footnote{ Investigation of the
background spectrum  shows that it has
  a soft  component likely originating from a dust-scattering halo around
the X-ray bright  RCW~103. } at lower energies
(the background contribution decreases
from 46\% in 0.5--8 keV  to 36\% in the 2--8 keV band).
 We find that
the PL model provides a good description of the extracted spectra
for both the inner and outer PWN.
  The
photon indices  are $\Gamma_{\rm inner}=1.47\pm0.24$ and
$\Gamma_{\rm outer}=1.65\pm20$
(see Fig.\ 9). Thus, the PWN
  spectra are noticeably softer
than the pulsar spectrum. The combined unabsorbed luminosity
from the two PWN regions is $L_{\rm
pwn}\approx3.2\times10^{33}d_{6.5}^2$ ergs s$^{-1}$ in the 0.5--8
keV band (see Table 3 for details).

\section{Discussion}

 In our high-resolution
{\sl Chandra} ACIS observation, we discovered the X-ray PWN around
the young, energetic pulsar J1617.
 Below we discuss the  pulsar and PWN properties
compare them with those of other young pulsars,
and discuss the connection between J1617 and the nearby HESS\,J1616.

\subsection{Pulsar}

 The
X-ray luminosity of J1617 is
  $\sim15$\%  lower than that reported by BA02 from the
{\sl XMM-Newton} observations. The discrepancy is not surprising
  because the broader
 PSF of {\sl XMM-Newton}  includes both
the pulsar and PWN contributions. Once these components are
added together,
 the {\sl Chandra } and {\sl XMM-Newton}
  flux (and spectral slope) measurements are in
   agreement within their uncertainties.
 The
44\% pulsed fraction measured from
the CC-mode data
is
somewhat lower than that
 reported by BA02. The unsubtracted contribution from the inner PWN,
 resolved in
 the deconvolved images (see Figs.\ 3 and 4),
 is
$\approx 10\%$--15\%, resulting
  in the
intrinsic pulsar's  pulsed fraction of $\approx50\%$.

 The
X-ray efficiency of the J1617 pulsar, $\eta_{\rm psr}\equiv L_{\rm
psr}/\dot{E}\approx1.1\times 10^{-3}d_{6.5}^2$ in the 0.5--8 keV
band, is typical among young pulsars (see Fig.\ 10, {\em top}).
However, the spectrum of J1617
is surprisingly hard,
$\Gamma_{\rm psr}\approx1.1$.
 Very few
pulsars (J1509--58, J1420--6048, J1846--0258)
 possibly exhibit similarly hard X-ray spectra (KP08 and references therein),
but the accuracies
  of the spectral slope measurements in
these pulsars are
lower than in J1617.
The fact that the  energy spectrum of the pulsar magnetospheric
X-ray emission
can be so hard
  has important implications for the pulsar emission models.
For instance, the models that interpret the non-thermal X-ray
emission as synchrotron emission from the particles produced in a
pair cascade predict $1.5<\Gamma <2$ (e.g., Cheng et al.\ 1998;
Crusius-W\"{a}tzel et al.\ 2001), i.e. softer than the spectrum we
measured.
On the other hand, the so-called ``full polar cap cascade'' models,
in which the
X-rays are produced by the resonant inverse
Compton scattering of the thermal emission from the hot NS
surface (Zhang \& Harding 2000), can explain harder spectra, with
$\Gamma_{\rm psr}\simeq 1$ (Fang \& Zhang 2006).
 Hard spectra ($\Gamma\simeq2/3$) are possible in the models
with
 curvature-radiation-induced cascades
in which curvature radiation may dominate synchrotron
radiation even below $\sim10$ keV (Rudak \& Dyks
 1999),
  although this emission mechanism
is more commonly considered
   for higher energies.  Finally, a hard
    spectrum could be produced
  via the resonant inverse Compton scattering of the radio photons  if this process contributes substantially to the X-ray band (Petrova 2008).
  Measurements
   at higher energies  with {\sl Integral} and {\sl RXTE}
  suggest
  a mild softening of the pulsar spectrum,
   by $\Delta\Gamma\approx0.2$--0.3
(see Fig.\ 5 in Landi et al. 2007). Such
 behavior is opposite to the spectral hardening observed
   in magnetars, which is  attributed to comptonization of
   the thermal emission from the NS surface by the relativistic electrons in the NS magnetosphere (Fern\'{a}ndez \& Thompson 2007; Baring \& Harding 2008).

  The lack of thermal component in the detected pulsar emission
might be
explained by its intrinsic faintness relative to the strong
magnetospheric emission in this young pulsar. However, it is hard to
constrain its contribution because, at $n_{\rm H,22}\approx 3.5$,
the soft thermal X-rays are strongly absorbed by the ISM.

\begin{figure}
 \centering
\includegraphics[width=3.2in,angle=0]{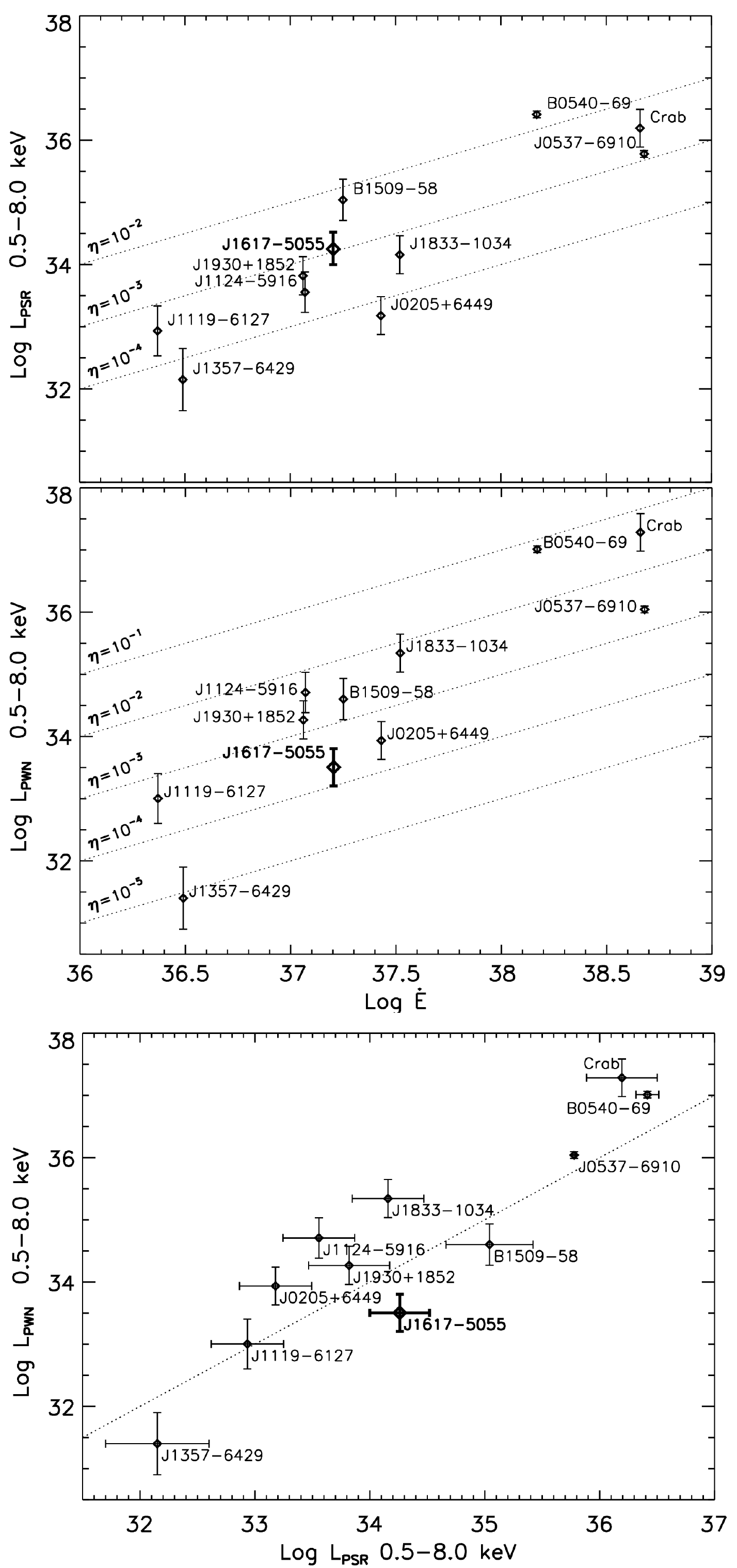}
\caption{{\em Top}: Pulsar luminosity vs.\
spin-down power $\dot{E}$. {\em Middle}: PWN luminosity vs.\ $\dot{E}$.
 {\em Bottom}: Pulsar luminosity vs.\ PWN luminosity.
The top and middle panels  show the lines of constant X-ray efficiency,
while the
 dashed line in the bottom panel corresponds to $L_{\rm psr}=L_{\rm pwn}$.
(Based on the data published by KP08.)
}
\end{figure}

\subsection{PWN}

Unlike the pulsar, the X-ray efficiency  of the J1617 PWN (compact +
extended), $\eta_{\rm pwn}\equiv L_{\rm pwn}/\dot{E}
=2\times10^{-4}d_{6.5}^2$ in the 0.5--8 keV band,  is noticeably
lower
 then those of other young
PWNe (Fig.\ 10, {\em middle}),
 except for J1357$-$6429 (Zavlin 2007).
Also, the ratio of the PWN to pulsar luminosities,
$L_{\rm pwn}/L_{\rm psr}\approx 0.18$ (independent of the poorly
known distance),
 is among the lowest (Fig.\ 10, {\em bottom}),
 and it is significantly lower than the average $L_{\rm pwn}/L_{\rm psr}\sim 4$
reported by Kargaltsev
et al.\ (2007a) and KP08.
The
actual PWN
efficiency and the $L_{\rm pwn}/L_{\rm psr}$ value
could be higher for J1617 (and J1357$-$6429) if the compact PWN
component is not completely resolved from the pulsar.
  An upper limit on the luminosity of the unresolved PWN core follows from the timing analysis (see \S2.3 and \S3.1).
   The luminosity of the unresolved core cannot
exceed $\approx50$\% of the pulsar luminosity (assuming that the
pulsar emission is 100\% pulsed),
 which still leaves the J1617 PWN underluminous compared to
majority of PWNe (see
  Fig.\ 10 and the discussion below).

KP08 found that the X-ray efficiencies of PWNe
show a very large scatter,
$\eta_{\rm pwn}
\sim  10^{-5}$--$10^{-1}$.
  The cause of such a large scatter is still
not understood.
One could attribute the
scatter to
  environmental differences (e.g., ISM  pressure and density); however,
  the surprisingly
  good correlation
between the PWN luminosities and non-thermal pulsar luminosities
    (with a much
smaller scatter, $0.1\lesssim L_{\rm pwn}/L_{\rm psr}\lesssim10$; KP08)
calls for a different explanation
because $L_{\rm psr}$ does not depend on the environment.

The deconvolved {\sl Chandra}
 images of J1617 suggest a compact structure at a distance
of $\sim 1''$ ($=9.7\times 10^{16}d_{6.5}$ cm) from
  the pulsar (see Fig.\ 4), which could be a torus
beyond the termination shock in the pulsar wind.
Assuming that the ram pressure
caused by the pulsar motion is lower
than the ambient pressure, we can
estimate the latter  as
  \be
  p_{\rm amb}\sim \dot{E} f_\Omega (4\pi c r_s^2)^{-1}
 \ee
where $r_s$ is the termination shock radius in the equatorial plane,
and the factor $f_\Omega$  takes into account possible anisotropy
of the pulsar wind.
For J1617,
this estimate gives
 $p_{\rm amb}\sim
1.7\times 10^{-8}f_\Omega(r_s/5\times10^{16}~{\rm cm})^{-2}$ ergs
cm$^{-3}$.
It is unlikely that the X-ray PWN is  more compact than it follows
from our image analysis (which would imply that the true PWN
luminosity and its ratio to $L_{\rm psr}$ are higher than estimated
above) because the inferred pressure
 is already quite high (cf.\ Table 2 in Kargaltsev et al.\ 2007a).
Therefore, we suggest  that in some PWNe, including J1617,
the radiative efficiency of the shocked pulsar wind is well below
the average.
The low PWN luminosity
cannot be attributed  to inefficient
 pair production in the pulsar magnetosphere
because the non-thermal pulsar luminosity is close to its typical
value.
However, the faintness of the PWN could be attributed to inefficient
particle acceleration on the way to (or at) the termination shock.
The physical mechanism
  responsible for  particle acceleration
is currently unknown,
 and the processes occurring  upstream of the shock are poorly understood
 (Arons 2007; Kirk et al.\ 2007).
  We can only speculate that these processes
 may depend of the angle between the
pulsar magnetic and rotation axes (e.g., a larger misalignment of the
axes
 may  lead to faster
conversion of the magnetic field energy to the particle energy
via more efficient
reconnection).
They may also depend
on the termination shock radius, which, in turn, depends on the
ambient pressure.
For instance, if $r_s$ is small,
 there may not be enough time for the dissipative processes
to convert the magnetic field energy to the particle energy
(e.g., Arons 2008).
Yet, even in this case, most
   of the pulsar's
rotational energy is being lost as a wind,
but the wind may have a higher magnetization parameter and lower
particle energies
(see,
 e.g., Arons 2007 for a pulsar wind theory review)
   than in the case of X-ray bright PWNe.
A  higher-than-typical  ambient pressure (hence a small termination
shock radius) is plausible
because the young pulsar should be
within the host SNR interior (which
  remains undetected
because of the strong X-ray absorption). The
wind still should emit synchrotron radiation
   downstream of the termination shock
 but possibly at lower frequencies
as the characteristic
   synchrotron frequency
is proportional to the  electron Lorentz factor squared.
Unfortunately, observing at lower X-ray energies or in the optical is
impossible
because of the strong ISM absorption, but sensitive far-IR and radio
observations would be useful.

\subsection{Connection to HESS~J1616--508}

 J1617 is located
$\sim 10'$ east of the TeV source HESS\,J1616--508 (see Fig.\ 5).
Despite the large offset,
 the young, energetic pulsar is still
within the TeV source extent,
and it has been considered as a plausible candidate for powering
the TeV emission (Aharonian et al.\ 2006; Landi et al.\ 2007), mainly
based on energetics arguments ($L_{\rm TeV}\sim
10^{-2}\dot{E}$).
Extended TeV sources have recently been discovered in the vicinity
of a few ($\approx15$)
   young pulsars, leading to the suggestion that
  the TeV emission comes from {\em relic PWNe}
crushed  by the asymmetric SNR reverse shock that arrived
  to the pulsar location
   a few  kyrs ago (e.g., de Jager 2006; Kargaltsev et al.\ 2007b;
Kargaltsev \& Pavlov 2007).
   The original
   model developed by  Blondin et al.\ (2001)
suggests that the  synchrotron emitting electrons  from the  pulsar
wind
   can be  swept up by the asymmetric reverse shock to one side of the pulsar.
The resulting ``offset PWNe'' will contain aged electrons
   that still can produce observable radio synchrotron emission as
well as  high-energy $\gamma$-ray emission  via the inverse Compton
scattering
   of background photons by relativistic electrons.

The high-resolution ACIS observations of J1617 do not reveal a
preferential extension of the J1617 PWN towards the TeV source
(rather there is a hint of extension in the opposite direction; see
\S2.1 and Fig.~1), such as observed
 in some other TeV sources neighboring
young pulsars  (Kargaltsev \& Pavlov 2007; Pavlov et al.\ 2008).
 Therefore, we conclude that there is
  no convincing evidence linking the TeV  source to J1617.

The HESS\,J1616 field has been observed with
several X-ray
missions.
The {\sl XMM-Newton} and {\sl Suzaku} images do not reveal
    any X-ray sources
    within the TeV bright part of HESS\,J1616,
making it one of the best examples of a  ``dark accelerator''
(Matsumoto et al.\ 2007).
  Yet, the
   examination of the
30 ks
    ACIS image covering the central part of
    the  HESS~J1616 field,
    reveals the possibly extended source X
(see Fig.\ 5 and \S2.2 ).
This source, however,
    has a very low X-ray flux  resulting in a surprisingly small
$L_{X}/L_{\rm TeV}=(1-2)\times10^{-3}$ (cf.\ Table 2 in Kargaltsev et al.\
2007b). Furthermore, the radio
    and IR images shown in Figure 5 reveal some diffuse emission
and pointlike sources
within the central part of HESS\,J1616.  The diffuse emission
     resembles a shell,
 while one of the bright IR pointlike sources coincides with the center
of the TeV source.
 Therefore, we speculate that HESS\,J1616 may
   consist of two (or more) unresolved sources,
 one of which still might be associated with J1716, while the other(s)
could be associated with the extended radio/IR emission (possibly
an SNR or a star-forming region).
Deeper radio and X-ray observations or TeV observations with
higher resolution can test
 this hypothesis.

\section{Summary}

 We have discovered a surprisingly underluminous PWN around the young,
  energetic pulsar J1617. The PWN consists of a compact and extended
components
 of comparable luminosities. The faintness of the PWN
could be attributed to the intrinsically low radiative efficiency of the
pulsar wind.  We
 have also obtained an
  accurate measurement of the pulsar's X-ray spectrum,
 which fits an
absorbed PL with $\Gamma_{\rm psr}\approx1.1$,
harder than predicted by the models of synchrotron emission
 from relativistic electrons/positrons produced
 in the pair-cascade.

J1617 has been  considered as a plausible counterpart to the offset
TeV source HESS\,J1616; however, the X-ray PWN does not
 show any  asymmetry towards  HESS\,J1616. 
  On the other hand, the archival X-ray, IR, and radio data
that cover  the central region of
   HESS\,J1616 reveal some diffuse emission,
  allowing for alternative interpretations of the TeV emission.

\acknowledgements
Support for this work was provided by the National Aeronautics and
Space Administration through Chandra Award Number GO6-7051X
 issued by the Chandra X-ray Observatory Center,
which is operated by the Smithsonian Astrophysical Observatory for
and on behalf of the National Aeronautics Space Administration
under contract NAS8-03060

\end{document}